# Synthesis and Characterization of Tungstite (WO$_3$.H$_2$O) Nanoleaves and Nanoribbons

*Majid Ahmadi[1] and Maxime J-F Guinel[1,2†]*

[1]Department of Physics, College of Natural Sciences, University of Puerto Rico, PO Box 70377, San Juan, PR 00936-8377, USA

[2]Department of Chemistry, College of Natural Sciences, University of Puerto Rico, PO Box 70377, San Juan, PR 00936-8377, USA

## Abstract

An environmentally benign method capable of producing large quantities of materials was used to synthesize tungstite (WO$_3$.H$_2$O) leaf-shaped nanoplatelets (LNPs) and nanoribbons (NRs). These materials were simply obtained by aging of colloidal solutions prepared by adding hydrochloric acid (HCl) to dilute sodium tungstate solutions (Na$_2$WO$_4$.2H$_2$O) at a temperature of 5-10$^o$C. The aging medium and the pH of the precursor solutions were also investigated. Crystallization and growth occurred by Ostwald ripening during the aging of the colloidal solutions at ambient temperature for 24 to 48hrs. When dispersed in water, the LNPs and NRs take many days to settle, which is a clear advantage for some applications (*e.g.,* photocatalysis). The materials were characterized using scanning and transmission electron microscopy, Raman and UV/Vis spectroscopies. The current versus voltage characteristics of the tungstite NRs showed that the material behaved as a Schottky diode with a breakdown electric field of 3.0x10$^5$V.m$^{-1}$. They can also be heat treated at relatively low temperatures (300$^o$C) to form tungsten oxide (WO$_3$) NRs and be used as photoanodes for photoelectrochemical water splitting.

---

[†] Corresponding author: e-mail address: *maxime.guinel@upr.edu*

# 1. Introduction

The synthesis of two dimensional (2D) inorganic nanomaterials has attracted considerable attention because they offer new opportunities in a multitude of applications. There exist several approaches to produce nanoscale materials: vapor-liquid-solid growth [1], solvothermal route [2], Langmuir-Blodgett technique [3], template method [4-5], electrospinning [6], and colloidal chemistry [7]. The most common methods to synthesize tungstite/tungsten oxide are electrochemical [8], solution-based colloidal [9], bioligation [10-11], chemical vapor deposition [12-13], and hydrothermal (HT) [14-15]. A simple heat-treatment of tungstite ($WO_3.H_2O$) allows for the phase transformation to tungsten oxide ($WO_3$), an important class of n-type semiconductors with a tunable band gap of 2.5-2.8 eV [16]. Moreover, its high chemical stability, low production costs and non-toxicity have recently generated significant interests for a wide variety of applications in microelectronics and optoelectronics [17-18], super-hydrophilic thin films [15], dye-sensitized solar cells [19], colloidal quantum dot LEDs [20], photocatalysis [21] and photoelectrocatalysis [22], water splitting photocatalyst as main catalyst [23-34]. Environmental applications can also benefit from $WO_3$ as a visible light photocatalyst to generate OH radicals for bacteria destruction [35] and photocatalytic reduction of $CO_2$ into hydrocarbon fuels [36].

The production of leaf-like $WO_3$ using a metallic tungsten surface exposed to a laser irradiation followed by an aging process has already been reported [37]. However, to the best of our knowledge, this is the first time that a very simple, inexpensive and environmentally benign synthesis of tungstite leaf-shaped nanoplatelets (LNPs) and nanoribbons (NRs) using colloidal chemistry is reported. Two steps are necessary to obtain these tungstite materials: The tungstate ions from sodium tungstate solutions ($Na_2WO_4.2H_2O$) are first protonated and in a second step, dimerization and crystallization occur. The materials obtained were characterized using scanning electron microscopy (SEM), high resolution transmission electron microscopy (HRTEM), UV/Vis, Raman and dynamic light scattering (DLS) spectroscopies. Moreover, the electrical properties of these materials were tested.

# 2. Experimental section

The tungstite ($WO_3.H_2O$) products were synthesized using the acid precipitation method [38]. Around 30-50mL of 6N hydrochloric acid (HCl) was added drop wise to a 100mL 15mM sodium tungstate solution ($Na_2WO_4.2H_2O$) while the solution was kept at 5-10$^o$C and under constant stirring. Ultrapure 18MΩ Millipore® deionized water was used for the preparation of all solutions.

These solutions were centrifuged and washed to reach pH~6 and added to abundant aqueous solutions (200-300mL) for crystallization and aging for durations of 24-48 hours at room temperature (RT), under constant stirring. The solutions were ultrasonicated for 2hrs before aging process. The effects of urea ($CH_4N_2O$), thiourea ($CH_4N_2S$), sulfuric acid ($H_2SO_4$) and HCl as pH adjusters and two chelating agents ethylenediaminetetraacetic acid (EDTA) and oxalic acid ($H_2C_2O_4$) were investigated.

All tungstite products were dried at 60°C in an oven in ambient air for 10-12hrs before characterization using Raman spectroscopy (Horiba Jobin-Yvon T64000 micro-Raman system, with an excitation of 532nm wavelength from a diode laser with a 0.2mW power), field emission scanning electron microscopy (SEM, JEOL JSM-7500F) and high resolution TEM (HRTEM, JEOL JEM-2200FS, operated at 200kV). X-ray energy-dispersive spectrometry (XEDS) in the SEM and the HRTEM was used to determine the elemental composition of the samples. UV/Vis absorption spectra in aqueous solutions were recorded using a Varian Cary 500 spectrophotometer. DLS measurements were carried out using Malvern Zetasizer Nano ZS90 particle size analyzer.

In order to record the current-voltage (I-V) and current-time (I-t) characteristics of these materials, drops of the NRs suspended in water were placed and allowed to dry on interdigitated Pt/PtO electrode devices. The electrode separation distance was 20μm with an electrode finger length of 1mm. A picture of the actual device is displayed in *ESI (electronic supplementary information), Figure S1,* as inset. A high internal resistance Keithley 2401 electrometer was used to record the signal. Because of these materials' sensitivity to light, all measurements were performed in the darkness.

## 3. Results

SEM images that were recorded from six different samples are displayed in Figure 1. Images *(a), (b)* and *(c)* show very thin NRs aged in deionized water (no additives), urea and thiourea for 24 hours at RT, respectively. Images *(d), (e)* and *(f)* show LNPs aged in deionized water (no additives), urea and thiourea for 24 hours at RT, respectively. XEDS recorded in the SEM and the TEM showed that only O and W elements were present in all samples (i.e., they were pure). The approximate sizes of the tungstite NRs were measured from SEM images: they were less than 1μm long, 150nm to 550nm wide and less than 20nm thick. DLS measurements for tungstite NRs are shown in *ESI, Figure S2* and were in agreement with an average length of about 600nm. When suspended in water, the NRs and LNPs took many days to settle to the bottom of containers. *Figure S2* shows two photographs of a suspension of NRs dispersed in water taken 24hrs and 48hrs after a 5min ultrasonication. A short movie showing how these NRs behave and scatter light when set in motion in water is provided in *ESI*.

TEM images and selected area electron diffraction (SAED) patterns for several tungstite NRs and one single NR are shown in Figure 2. The diffraction patterns (DPs) were indexed to the orthorhombic *Pmnb* structure of tungstite with lattice parameters a=0.52, b=1.07 and c=0.51 nm, which was in good agreement with published data ($WO_3.H_2O$ x-ray diffraction card number JCPDS No.43-0679 and the American Mineralogist Crystal Structure Database, AMCSD 0005199 [39]). Also two TEM and two HRTEM images recorded from NRs are shown in Figure 3*(a-d)*. A schematic of one unit cell of tungstite consisting of four distorted octahedra, in accordance with SAED results is displayed in Figure 3*(e)*. Electron energy-loss spectra (EELS) were recorded (using the in-column energy filter fitted in the HRTEM) around the O K-edge (532.0 eV). A typical spectrum is displayed in Figure 3*(f)*. Multiple scattering effects can be ignored because the NRs were less than 20nm thick, therefore having little consequence on the energy loss near-edge structure (ELNES). The O K-edge may be influenced by several factors including the ejection of electrons into the continuum producing a saw-tooth shape, as well as solid state effects and transitions to unoccupied bound states creating fine structures in the edge [40-41]. The ELNES of the O K-edge in transition metal oxides is due to the crystal field splitting of the metallic orbitals into $e_g$ and $t_{2g}$ components, resulting in the EELS double-peak signature of coordinated oxygen atoms as clearly seen in the spectrum shown in Figure 3(f). It was also reported for some other oxides such as ceria, titania, rare earth oxides and zirconia [42-44].

Raman and UV/Vis absorption spectra recorded from tungstite NRs synthesized in water (no additive) and in the presence of urea and thiourea are shown in Figure 4*(a & b)*.

The I-V characteristics in the range -6V to +7V are shown in the plot of Figure 5. For ease of description, the plot was divided into five zones. The first three zones (I, II and III) correspond to the reverse bias voltage and the following two zones (IV and V) to the forward bias voltage due to the Schottky barrier behavior of this metal/semiconductor ($Pt/WO_3.H_2O$) connection. The first zone (-6V to -5V) showed negative resistance. The third and fourth regions (-2 to zero and zero to +2V) showed near ohmic properties. The fifth region (>+2V) showed lower resistance and the current increased rapidly when the applied voltage was increased. These measurements were repeated several times using different ranges of applied potentials and the same behavior was always observed.

## 4. Discussion

Unlike nanoparticles that tend to agglomerate relatively quickly, especially when no dispersing agents are used, the morphology of these NRs (or LNPs) brings a very clear advantage for catalysis and photocatalysis applications, because they can remain suspended for very long times without any external force (*ESI, inset of Figure S2*).

The crystallization and growth of tungstite crystals follow traditional self-aggregation and Ostwald ripening mechanisms. Urea is a suitable ligand or directing agent because it can act as either a hydrogen-bond donor through its two N-H bonds or a hydrogen-bond acceptor through the lone pairs of the C=O group [26]. Urea has been used to assemble nanoporous tungsten oxide nanotubes [45]. The SEM data did not show observable changes in the morphology of materials prepared using either urea or thiourea (See Figure 1). When the pH of the precursor solutions was decreased from 6.0 to 3.0 using either $H_2SO_4$ or HCl, the products had a tendency to be amorphous with some crystalline LNPs or NRs (*See ESI, Figure S3*). When the pH of the aging solution was below three (*e.g.,* 1.0 or 2.0), the products were amorphous.

SAED results confirmed that the NRs had the orthorhombic structure which consists of distorted octahedral units of tungsten atoms coordinated with five oxygen atoms and a water molecule where the octahedra share four vertices in the equatorial plane forming the sheet structure. The sheets are held together by hydrogen-bonding between the water molecules and the oxygen atoms in the axial position (*z* direction) in the adjacent layers. The tungsten to oxygen bond lengths are: 0.169nm (W=O), two 0.183nm (W-O), two 0.193nm (W-O) and 0.234nm (axial, waterW-OH$_2$) [39].

The reaction in our synthesis method [46] occurs in two major steps, the first step being the protonation of the tungstate ions upon solution acidification to form white solid precipitates:

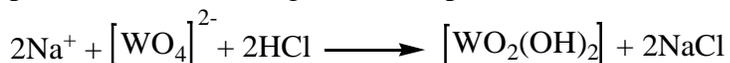

$$2Na^+ + \left[WO_4\right]^{2-} + 2HCl \longrightarrow \left[WO_2(OH)_2\right] + 2NaCl$$

The second step is the hydration of the [WO$_2$(OH)$_2$] tetrahedral molecules and dimerization via O-bridging to form crystalline [WO(OH)$_3$(H$_2$O)]$_2$(μ-O) containing octahedral W-centers:

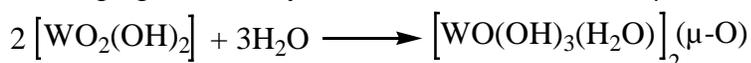

$$2\left[WO_2(OH)_2\right] + 3H_2O \longrightarrow \left[WO(OH)_3(H_2O)\right]_2 (μ\text{-}O)$$

Two proposed schemes for the crystallization and growth of 2D tungstite LNPs or NRs are provided in *ESI, Figure S4*. SAED confirmed this crystalline structure, composed of the same ReO$_3$-type octahedra layers.

HRTEM measurements of lattice planes (e.g., those shown in Figure 3) demonstrated that the growth direction for the NRs was along the <111> directions. Therefore, the oxolation along equivalent *x* and *y* directions ({111} lattice planes) during the crystallization and growth steps lead to the formation of a tungstite crystalline layered phase. SAED patterns showed a zone axis (beam direction) in the <110> direction for NR observed in their planar direction. A schematic of one unit cell of tungstite consisting of four distorted octahedra, in accordance with SAED and proposed schemes, is displayed in Figure 3*(e)*.

EELS (e.g., in Figure 3*(f)*) showed that the peak separation (difference between the first and the second peak) in the O K-edge was about 7eV (corresponding to ~56,500 cm$^{-1}$) which corresponds to the e$_g$ and t$_{2g}$ orbital components. According to crystal field theory, the *5d* orbitals in tungstite are split into the low-energy orbital t$_{2g}$ (π-bond) and the higher energy orbital e$_g$ (σ-bond) resulting in this EELS double-peak, corresponding to the fingerprint of tetrahedrally coordinated oxygen atoms. Theoretically, the intensity ratio of t$_{2g}$ to e$_g$ is equal to 3/2 (*See ESI, Figure S5*). From our results, this ratio was indeed approximately 3/2, following the Pearson method [47].

In Raman studies, the peaks at 635cm$^{-1}$ and 943cm$^{-1}$ were assigned to the a$_g$ phonons of the WO$_3$.H$_2$O lattice, in agreement with reference [48]. The vibration frequency of the W=O bond occurs at higher frequency (943cm$^{-1}$) than that of the W-O bond (635cm$^{-1}$) because it is a stronger bond. For transition metal (M) oxides, the peaks in the range 1050-950cm$^{-1}$ can be assigned to the stretching mode of short terminal M=O bonds. For tungstite, the W=O bond peak, typical of non-bridging O, is caused by structural water molecules. One of the axial O positions in the octhedron can be occupied by a structural water molecule and this O is associated with a single bond and the opposite axial O forms a strong W=O double bond (terminal bond). These two O bonds are associated with the 635cm$^{-1}$ and 943cm$^{-1}$ peaks, respectively [49-50]. The values for bending or deformation of the O-W-O equatorial bonds within the octahedra of crystalline WO$_3$ [51] are 267cm$^{-1}$ and 330cm$^{-1}$. Hence, two peaks in the range of 180-250cm$^{-1}$ were assigned to the bending or the deformation of O-W-O bonds in WO$_3$.H$_2$O.

UV/Vis spectroscopy showed two peaks with a separation of about 13,400cm$^{-1}$. These two peaks can be assigned to the $e_g^*$ ($\sigma^*$-bonds) and $t_{2g}^*$ ($\pi^*$-bonds) orbital components. Ignoring electron correlation effects and considering that the exciting electron from $a_{1g}$, $t_{2g}$ or $e_g$ to $t_{2g}^*$ is forbidden, these two peaks are the result of the electron excitation from $t_{1u}$ occupied orbitals to two unoccupied anti-bonding components $e_g^*$ and $t_{2g}^*$ corresponding to 245nm and 365nm photons (*See ESI, Figure S5*). A shift to the right (in the visible absorbance region) and an increase in light absorbance in the UV region were detected when urea and thiourea were used with the maximum absorbance occurring with thiourea. The shift and the improvement in absorbance may be attributed to slight changes in the size and morphology of the NRs synthesized which can affect charge-transfer transition in tungstite conduction or valence band. An increase in photon absorption was observed for NRs synthesized in water, urea and thiourea. A judicious choice for the aging solution may therefore improve the photoactivity of these materials.

The Schottky barrier behavior for the Pt/WO$_3$.H$_2$O connection was observed in I-V studies. Tungsten oxide is an n-type semiconductor, therefore in order to observe this Schottky behavior, the work function of Pt ($\Phi_{Pt} \approx 5.3$ eV) must be higher than the work function of tungstite (*See ESI, Figure S6*) [52]. The breakdown electric field value for tungstite was measured to $3.0 \times 10^5$V.m$^{-1}$. The first zone identified in the I-V plot can be considered as a discharging effect of the tungstite NRs. Yoon *et al.* reported supercapacitor properties and high electrical conductivity (1.76 S.cm$^{-1}$) for mesoporous tungsten oxide materials [53]. The current versus time (I-t) characteristics recorded from the tungstite NRs are typical of a discharge (*ESI, Figure S1*).

## 5. Conclusions

Tungstite leaf-shaped nanoplatelets and nanoribbons were synthesized for the first time using an easy, inexpensive and environmentally benign colloidal chemistry method which can readily be scaled up for commercial applications. They were very thin (about 20nm) and grew along the <111> directions. When dispersed in water, they take several days to settle to the bottom of a container which presents clear advantages for photocatalysis applications. It displays diode properties with an electrical breakdown at $3.0 \times 10^5$V.m$^{-1}$. A short heat treatment at relatively low temperatures allows for the phase transformation to tungsten oxide (WO$_3$) while maintaining the NRs morphology, the aim being using these materials as photoanodes for the photocatalysis of water [22 and 31].

## Acknowledgements

Partial support from the National Aeronautics Space Administration (NASA) grants NNX10AM80H and NNX07A030A. The U.S. National Science Foundation (NSF) for its support to the Nanoscopy Facility at UPR.

**Figure captions**

**Figure 1.** SEM images showing tungstite NRs obtained by aging using (a) deionized water, (b) urea at pH=7, (c) thiourea at pH=7 and tungstite LNPs obtained by aging using (d) deionized water, (e) urea at pH=7 and (f) thiourea at pH=7.

**Figure 2.** TEM images recorded from (a) several tungstite NRs and (c) from one single NR. Corresponding SAED patterns are shown in (b) and (d), respectively. They were indexed to polycrystalline and single crystal orthorhombic tungstite along the [110] zone axis.

**Figure 3.** (a, b) TEM and (c, d) HRTEM images recorded from tungstite NRs synthesized in water. The {111} lattice planes are indicated. Crystals grew along the <111> directions. (e) Schematic of one unit cell of tungstite. (f) EELS showing the O K-edge recorded from one tungstite NR.

**Figure 4.** (a) Raman spectra recorded from tungstite NRs synthesized in (I) water, (II) urea, (III) thiourea and (b) corresponding UV/Vis absorption spectra.

**Figure 5**. I-V plot recorded for tungstite NRs.

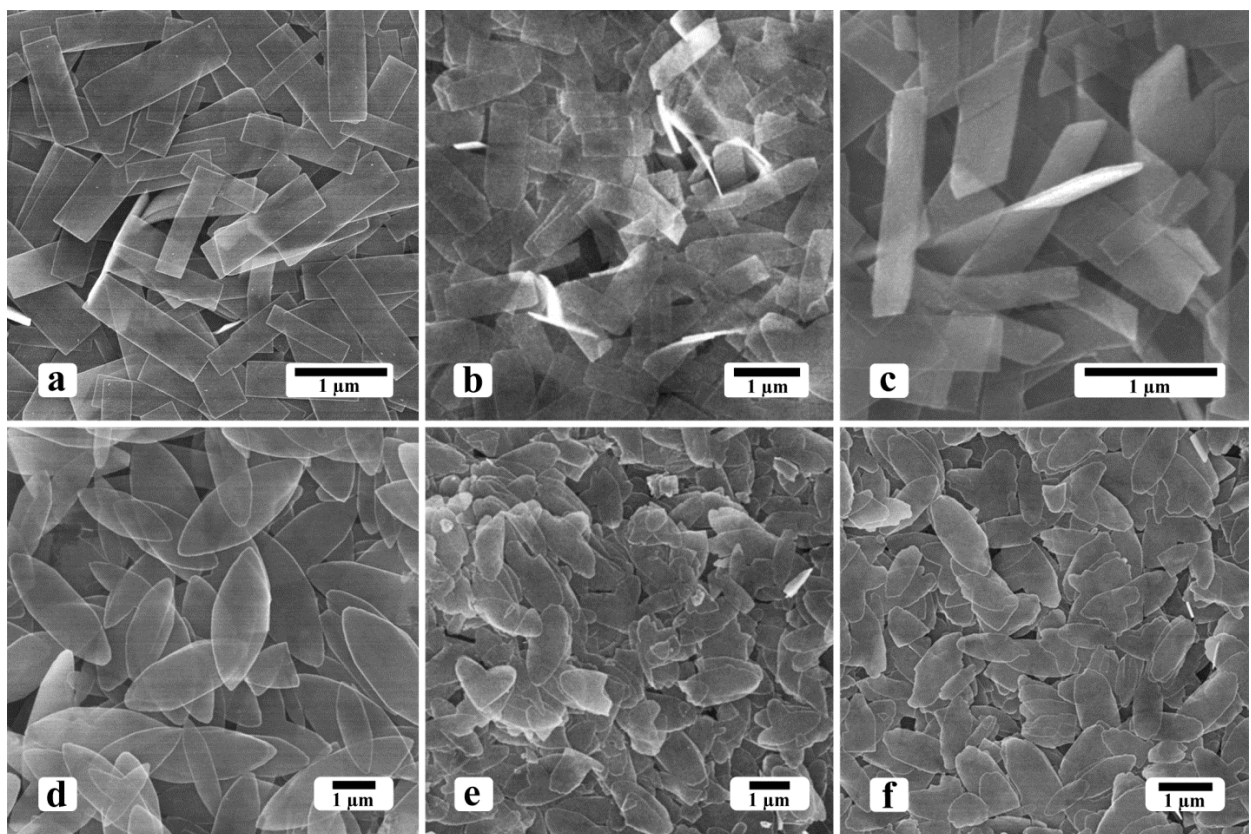

**Figure 1.**

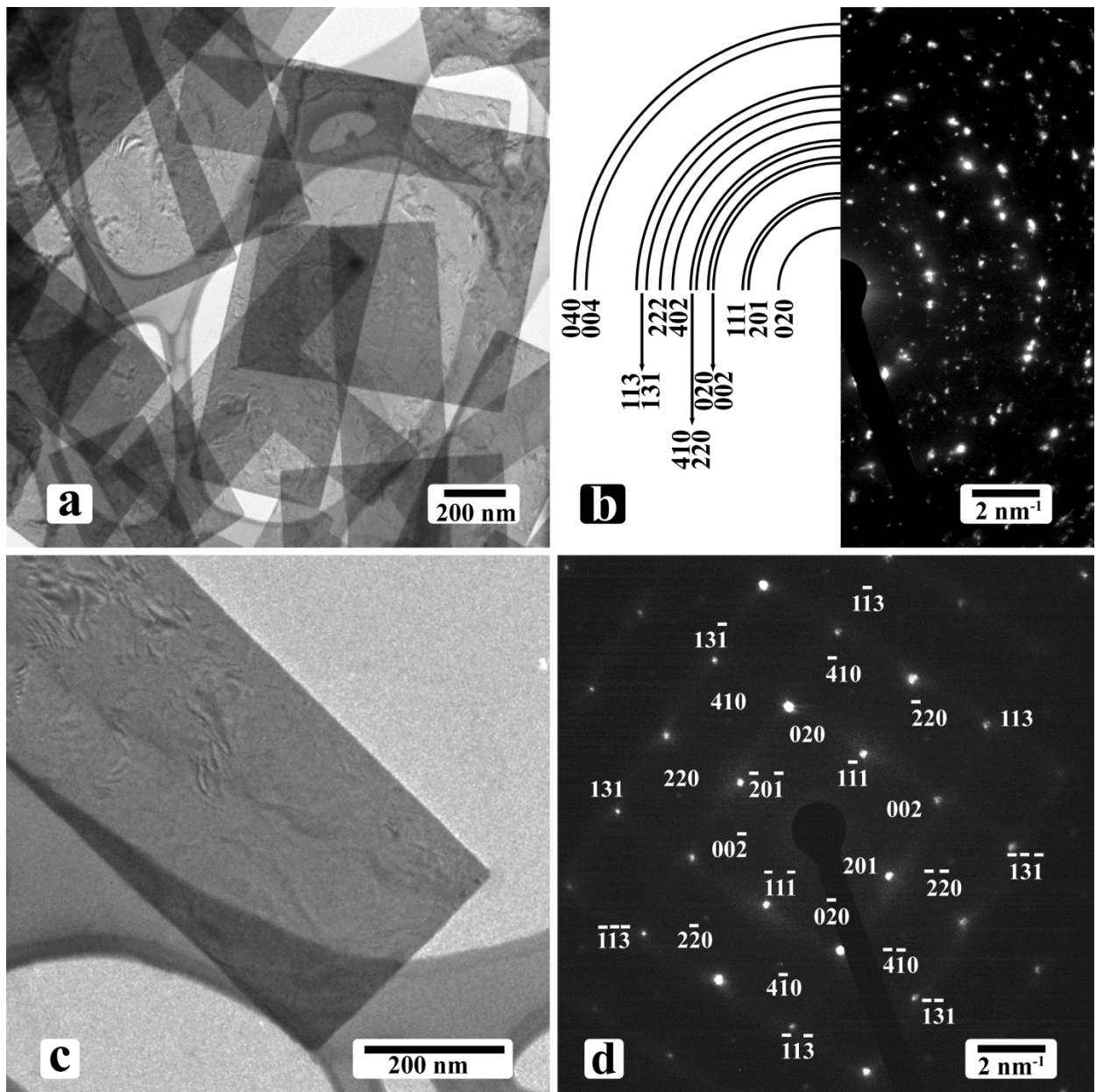

**Figure 2.**

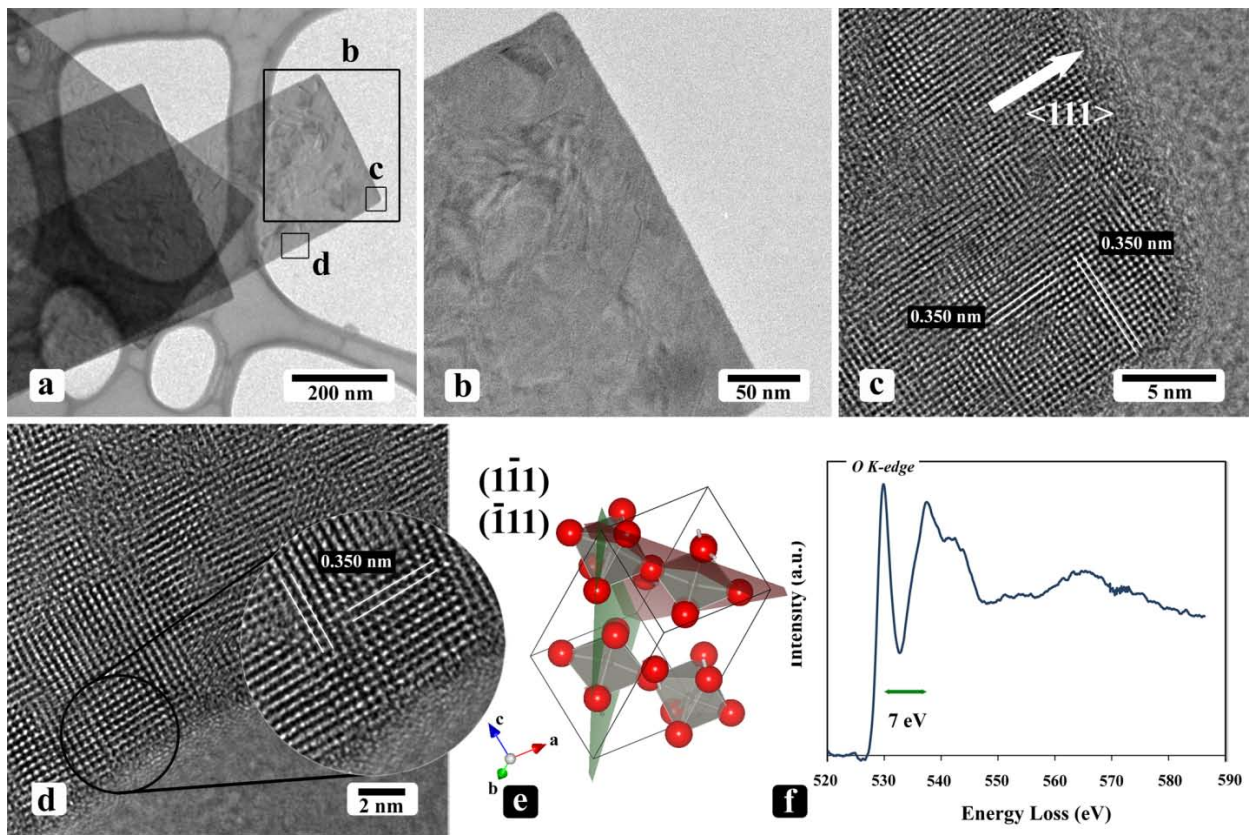

**Figure 3.**

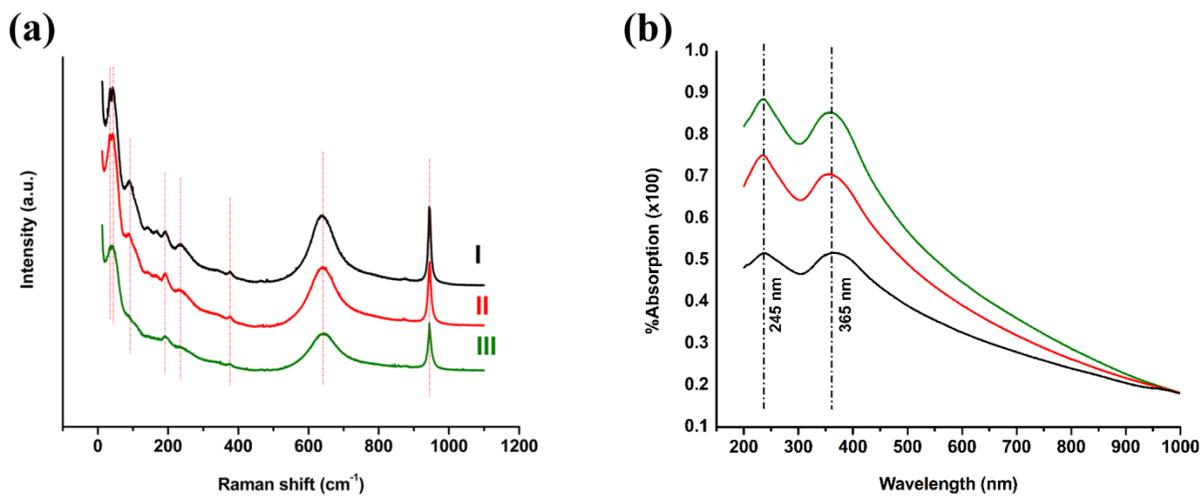

**Figure 4.**

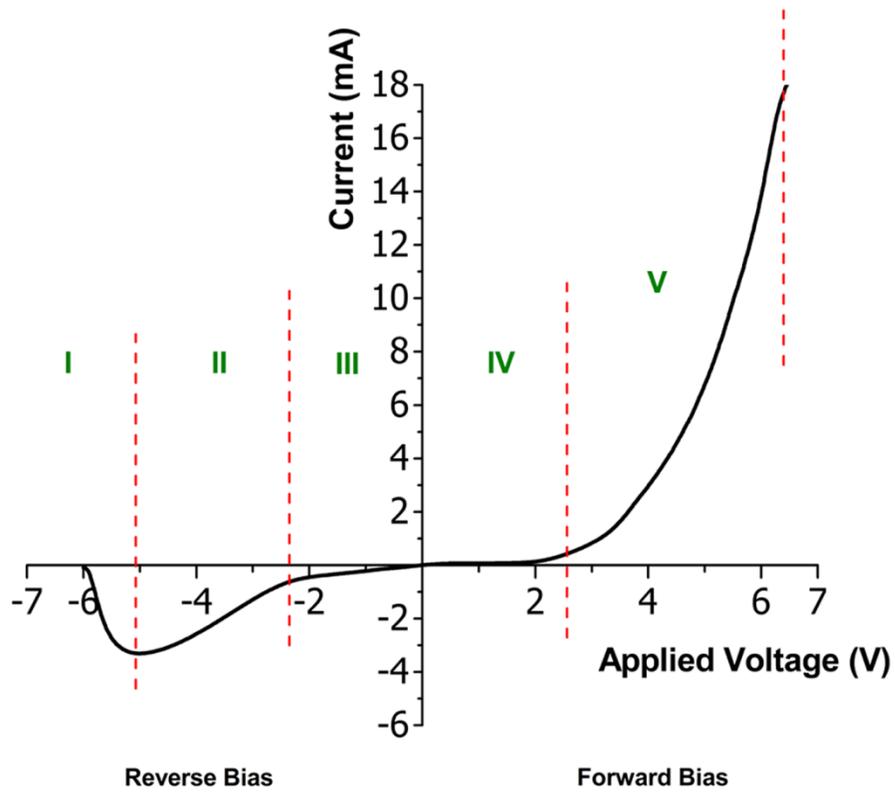

**Figure 5.**